\documentclass[letterpaper, 10 pt, conference]{ieeeconf}  

\IEEEoverridecommandlockouts                              

\usepackage[tight,footnotesize]{subfigure}

\usepackage[utf8]{inputenc}
\usepackage[T1]{fontenc}
\usepackage{textcomp}
\usepackage{siunitx}
\usepackage{graphicx}
\usepackage{epstopdf}
\usepackage{amsmath}
\usepackage{amssymb}
\usepackage{psfrag}
\usepackage{algpseudocode}
\usepackage[utf8]{inputenc}
\usepackage[T1]{fontenc}
\usepackage{multicol}
\usepackage{multirow}
\usepackage{textcomp}
\usepackage{graphicx}
\usepackage{epstopdf}
\usepackage{amssymb}
\usepackage{amsmath}
\usepackage{psfrag}
\usepackage[tight,footnotesize]{subfigure}
\usepackage{cite}
\usepackage{enumerate}
\usepackage{algorithm}
\usepackage{algpseudocode}
\usepackage{mathtools}
\usepackage{algcompatible}

\makeatletter
\newcommand{\vast}{\bBigg@{3.2}}
\newcommand{\Vast}{\bBigg@{5}}

\makeatother

\graphicspath{{Figure/}}

\begin{document}

\title{\LARGE \bf Generalized Iterative Super-Twisting Sliding Mode Control:\\ A Case Study on Flexure-Joint Dual-Drive H-Gantry Stage   
}

\author{Wenxin Wang$^{1,2}$, Jun Ma$^{2}$, Zilong Cheng$^{2}$, Xiaocong Li$^{1}$, Abdullah Al Mamun$^{1,2}$, and Tong Heng Lee$^{2}$
	\thanks{$^{1}$SIMTech-NUS Joint Lab on Precision Motion Systems, Department of Electrical and Computer Engineering, National University of Singapore, Singapore 117582. E-mail: wenxin.wang@u.nus.edu; li\_xiaocong@simtech.a-star.edu.sg; eleaam@nus.edu.sg.}%
	\thanks{$^{2}$Department of Electrical and Computer Engineering, National University of Singapore, Singapore 117583. E-mail: elemj@nus.edu.sg; zilongcheng@u.nus.edu; eleleeth@nus.edu.sg.}
    \thanks{This work has been submitted to the IEEE for possible publication. Copyright may be transferred without notice, after which this version may no longer be accessible.}
}

\maketitle

\begin{abstract}

Mechatronic systems are commonly used in the industry, where fast and accurate motion performance is always required to guarantee the manufacturing precision and efficiency. Nevertheless, the system model and parameters are difficult to be obtained accurately. Moreover, the high-order modes, strong coupling in multi-axis system, or unmodeled frictions will bring uncertain dynamics to the system. To overcome the above-mentioned issues and enhance the motion performance, this paper introduces a novel intelligent and totally model-free control method for mechatronic systems with unknown dynamics. In detail, a 2-degree-of-freedom (DOF) architecture is designed, which organically merges a generalized super-twisting algorithm with a unique iterative learning law. The controller solely utilizes the input-output data collected in iterations such that it works without any knowledge of the system parameters. The rigorous proof of convergence ability is given and a case study on flexture-joint dual-drive H-gantry stage is shown to validate the effectiveness of the proposed method.

\end{abstract}


\section{Introduction}

Mechatronic systems have been widely used in manufacturing industry, such as inspection process, transmission task, stepping and scanning device. As a popular application, timing-belt driven tray indexing is applied for industrial transmission tasks in such areas due to its strong transmission force and low cost~\cite{wang2020iterative1,ma2021maglev}. Nevertheless here, there are actually some high-order modes and other unmodeled dynamics which exert inevitable affects on the system performance~\cite{li2017data,wang2020iterative,chen2020novel}. Besides, after long-time use of the stage, the system parameters will change due to the wearing. Also, the problems of deadzone and backlash caused by frictions will notably influence the position tracking performance. Besides, multi-axis linear motor driven dual-drive H-gantry stage is also widely used for planar transmission tasks. Conventionally, the carriages and the cross-arm are linked by rigid joints~\cite{yuan2016time,hu2020precision}. However, in certain scenarios, the de-synchronization may occur and some inter-axis coupling force might be generated which will damages the joints~\cite{ma2019parameter}. To prevent this kind of unintended damage, flexure-joint linked gantry stage have been designed~\cite{xu2013design,zhu2016tmech,wu2018design,kang2020six}. However, this design will introduce more coupling and nonlinearity to the system, which naturally bring various uncertainties and disturbances during the planar motion tasks~\cite{ma2017integrated,kamaldin2018novel}. Therefore, as a result, the motion performance could be inevitably diminished in real-world applications due to the system uncertainties and unmodeled dynamics.

To guarantee the motion performance of mechatronic systems, the model-based optimal control methods are developed~\cite{chen2020optimal,ma2021optimal,ma2021convex}. Nevertheless, these model-based approaches require the accurate information of the model. Consequently, the research in the development of data-driven approaches has been proposed recently~\cite{hou2010novel,zhang2020trajectory}. For instance, data-driven iterative tuning approaches is proposed in \cite{li2020data} such that good position tracking performance is achieved using input-output data. however, the accurate system identification is still required for initialization. Therefore, the totally intelligent model-free controllers is a notable trend to the industrial practice.

Pertaining to the above descriptions, a generalized iterative super-twisting sliding mode control method is proposed in this paper. It is a 2-degree-of-freedom (DOF) architecture comprising a generalized super-twisting algorithm with a unique iterative learning law. Notably, the super-twisting algorithm makes the control input absolutely continuous~\cite{polyakov2009reaching,utkin2013convergence}. However, due to the discontinuous terms (signum functions) under the integrator in the super-twisting algorithm, the input chattering can be only attenuated but not fully removed. Moreover, similar to traditional sliding mode controller, the boundary of matched uncertainties should be known, but the accurate value of such a boundary is rather difficult to estimate. This inevitably leads to larger but unnecessary control gains which bring a great chattering. To overcome the above issues, a unique iterative learning law is designed to attenuate the uncertainties during iterations and thus the ``pressure" of the super-twisting sliding mode control can be consequently alleviated. Additionally, the sliding dynamics is also considered as iterates and then the sliding hyperplane is reached in both the time domain and the iteration domain. Moreover, the other merit of this proposed method is that it is applicable to multi-axis even though strong coupling exists (such as the above mentioned flexure-joint dual-drive H-gantry stage). Specifically, for each axis, the influence from other axes can be regarded as uncertainties and thus the controller of each axis could be designed independently. The main contributions of this work are listed: (1). A totally model-free intelligent control method is proposed. Meanwhile, the motion accuracy could be ensured at a high level versus the uncertainties. (2). The architecture combines the merits of each term organically. For the super-twisting algorithm, the sliding motion is established in the iteration domain. For the iterative learning control, the super-twisting algorithm improves the convergence of the iterative learning law by suppressing the non-repeated uncertainties. (3). This proposed method is applicable to multi-axis systems and the modeling of the coupled system is not required.  

The rest structure of this paper is summarized below. In Section II, the problem of a general uncertain system is addressed. Then, the overview of the control architecture is given in Section III with a rigorous proof of the convergence abilities. Section IV presents a case study on the flexure-joint dual-drive H-gantry stage to validate the effectiveness of the proposed methodology. Finally, conclusions are given in Section V.

\begin{figure}[t]
	\centerline{\includegraphics[trim = 0cm 0cm 0cm -1cm, width=\columnwidth]{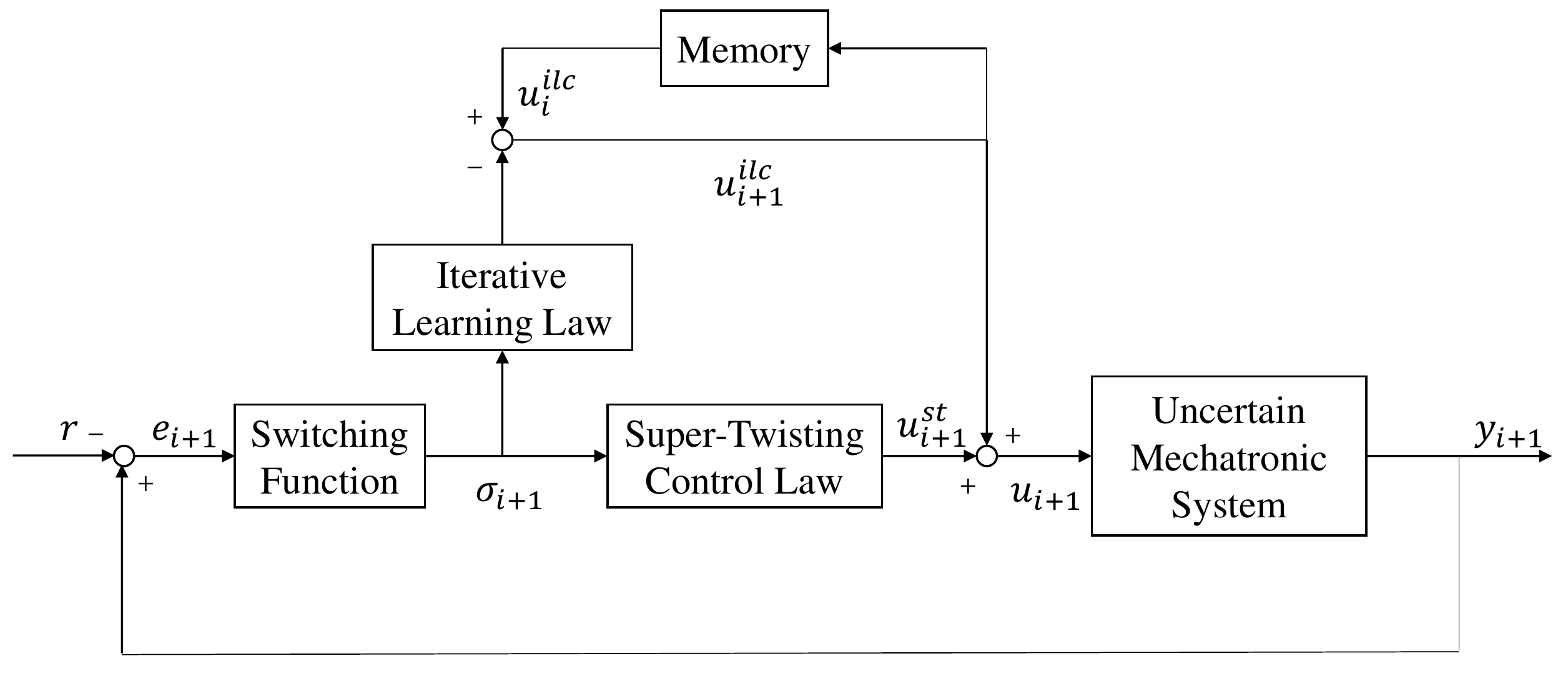}}
	\caption{2-DOF control architecture.}
	\label{fig.diagram}
\end{figure}

\section{Problem Formulation}

A general uncertain mechatronic system is written in the following form.
\begin{eqnarray}
\left\{    
\begin{array}{rcl}        
\dot{x} &=& f(x)+g(x)u \\        
y &=& Cx    
\end{array} \right.    \label{regular-form}
\end{eqnarray}
\noindent where $x$ is the state variable, $u$ denotes the control law, and $y$ denotes the system output, respectively. $f(x)$ and $g(x)$ are smooth uncertain functions satisfying global Lipschitz continuity condition. Note that, if the system describes an uncertain mechatronic system (such as tray indexing or one axis of dual-drive H-gantry stage), $y$ is the position, $x=\left[\begin{matrix} x_1 & x_2 \end{matrix}\right]^T$ where $x_1$ is the position and $x_2$ is the derivative of the position, and $C=\left[\begin{matrix} 1 & 0 \end{matrix}\right]$. The main target is to design a totally model-free controller for such an uncertain mechatronic system. At first, the sliding variable $s$ is designed as
\begin{equation}
\begin{aligned}
	s &= \lambda e+\dot{e}\\
	&= \left[\begin{matrix} \lambda & 1 \end{matrix}\right]\left[\begin{matrix} r-y \\ \frac{d}{dt}\big(r-y\big) \end{matrix}\right]
	\label{sigma} 
\end{aligned}
\end{equation}
\noindent where $r$ is a derivable position reference profile. For the stability and reachability of the sliding hyperplane, the parameter $\lambda$ should be designed as a suitable positive constant.

After that, the dynamics of sliding variable is derived
\begin{equation}
\begin{aligned}
	\dot{s}(t) &= \frac{\partial s}{\partial t}+\frac{\partial s}{\partial x}\cdot\dot{x}\\
	&= \underbrace{\frac{\partial s}{\partial t}+\frac{\partial s}{\partial x}\cdot f(x)+(\frac{\partial s}{\partial x}\cdot g(x)-1)u}+u \\
	& \quad \quad \quad \quad \quad\quad\quad\quad \Psi(t) 
	\label{sigma-dynamics1}
\end{aligned}
\end{equation}
Furthermore, \eqref{sigma-dynamics1} can be rewritten as
\begin{equation}
\dot{s} = \Psi(t) + u
\label{sigma-dynamics}
\end{equation}
\noindent where $\Psi(t)$ is an unknown function. Then it is important to make sure the establishment of the sliding motion and keep the sliding variable $s$ around sliding hyperplane by the control law $u$ without any prior information of unknown functions $f(x)$ and $g(x)$. 

\section{Controller Design}

\subsection{Overview of the 2-DOF Control Architecture}

As is shown in Fig. $\ref{fig.diagram}$, the 2-DOF control architecture comprises a generalized super-twisting algorithm with a signum-type iterative learning law. $u^{st}_{i}$ and $u^{ilc}_i$ denote the control input in the $i$-th iteration generated from super-twisting algorithm and iterative learning law, respectively. Specifically, the control law $u$ is designed as
\begin{equation}
u=u^{st}+u^{ilc}
\label{u}
\end{equation}
Here, the generalized super-twisting control method is described in the following form
\begin{IEEEeqnarray}{rcl}
	u^{st} &=& -k_1\varphi+k_2 v \label{ST1} \\   
	\dot{v} &=& -\varrho
	\label{ST2}
\end{IEEEeqnarray}
\noindent with 
\begin{IEEEeqnarray}{rCl}
	\varphi &=& \vert s\vert^{\frac{1}{2}}{\rm{sgn}}\big(s\big)+k_c s \label{varphi} \\
	\varrho &=& \frac{1}{2}{\rm{sgn}}\big(s\big)+\frac{3}{2}k_c\vert s\vert^{\frac{1}{2}}{\rm{sgn}}\big(s\big)+k_c^2s \nonumber \\
	&=& \frac{d}{ds}\varphi\cdot\varphi
	\label{varrho}
\end{IEEEeqnarray}
\noindent where the positive constants $k_1$, $k_2$ and non-negative constant $k_c$ all denote control gains.

The linear term of the super-twisting algorithm could suppress the uncertainties which grows together with the state linearly. This additional term makes the sliding manifold $s=0$ easier to be reached. Notice that, if the linear component control gain $k_c$ is set as 0, \eqref{ST1} and \eqref{ST2} is downgraded to a typical super-twisting sliding mode controller. The algorithm is continuous so that the uncertain function $\Psi$ could satisfy Lipschitz continuity condition. To solve the issue of unknown boundary of the uncertainties, a sigmun-type iterative learning law is designed to suppress the uncertainties during the iterations:
\begin{equation}
\hat{\Psi}_{i+1}(t)=\hat{\Psi}_i(t)+2q\beta\varrho_{i+1}(t)
\label{ilc1}
\end{equation}
\noindent where $\hat{\Psi}_i$ is estimation of the uncertain function $\Psi$, $q$ and $\beta$ are positive constants. According to \eqref{sigma-dynamics}, the principle of the iterative learning law can be actually described as
\begin{equation}
u^{ilc}_i=-\hat{\Psi}_i(t) 
\label{uILC}
\end{equation}
And thus,
\begin{equation}
u^{ilc}_{i+1}(t)=u^{ilc}_i(t)-2q\beta\varrho_{i+1}(t)
\label{ilc}
\end{equation}
Then, if we define $\tilde{\Psi}_i=\Psi_i-\hat{\Psi}_i$ as the discrepancy between the uncertain function and its estimation, \eqref{sigma-dynamics} can be written as
\begin{equation}
\dot{s}_i = \tilde{\Psi}_i+u_i^{st}
\label{sigma-dynamics3}
\end{equation}
\noindent which indicates that the uncertain function could be estimated better in the iteration domain if the iterative learning law is convergent.

\begin{figure}[t]
	\centerline{\includegraphics[trim = 0cm 0cm 0cm 0cm, width=8.5cm]{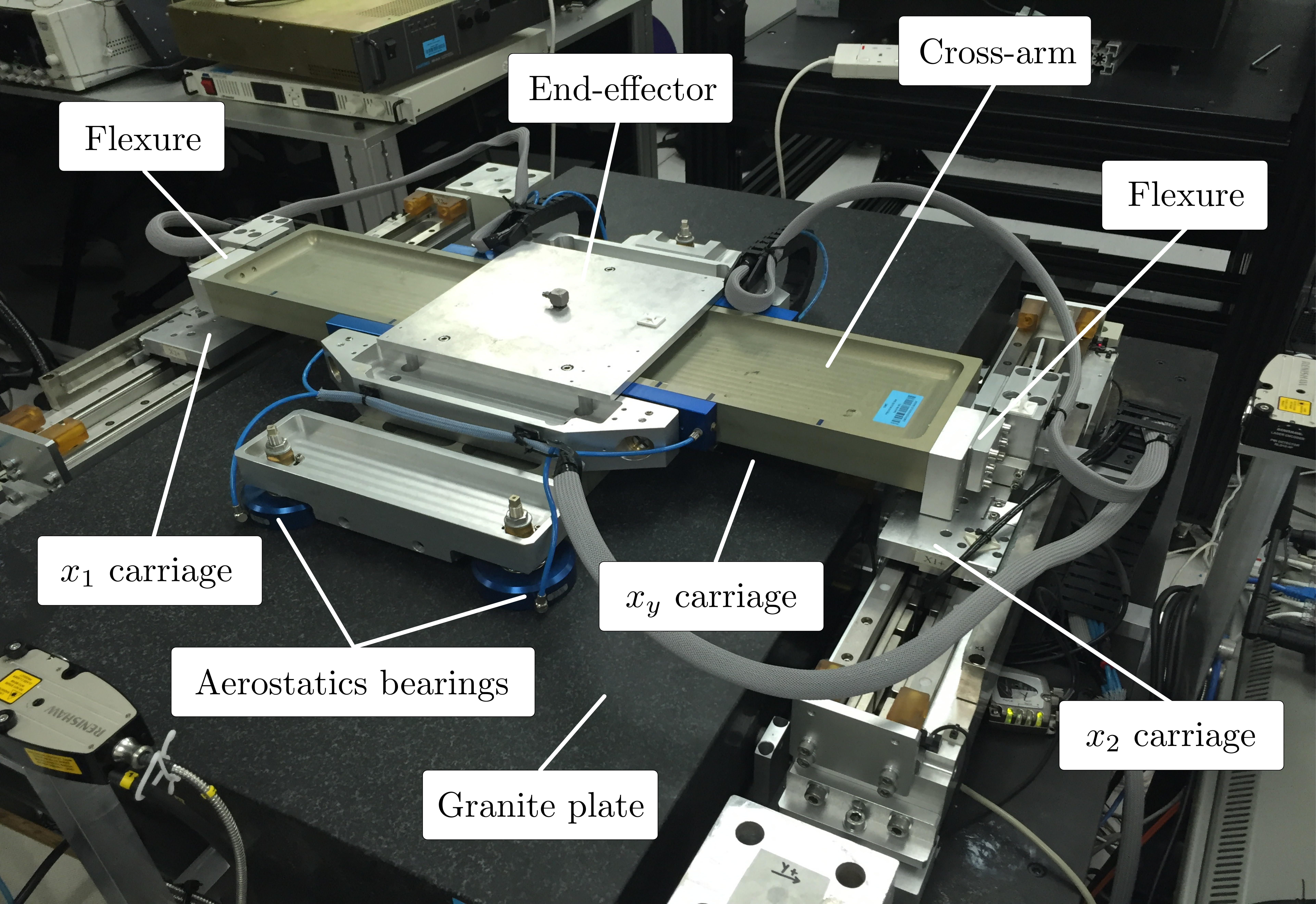}}
	\caption{Setup of a flexure-joint dual-drive H-gantry stage.}
	\label{fig.gantry}
\end{figure}

\begin{figure}[t]
	\centerline{\includegraphics[trim = 0cm 2cm 0.5cm 1.5cm, width=8.5cm]{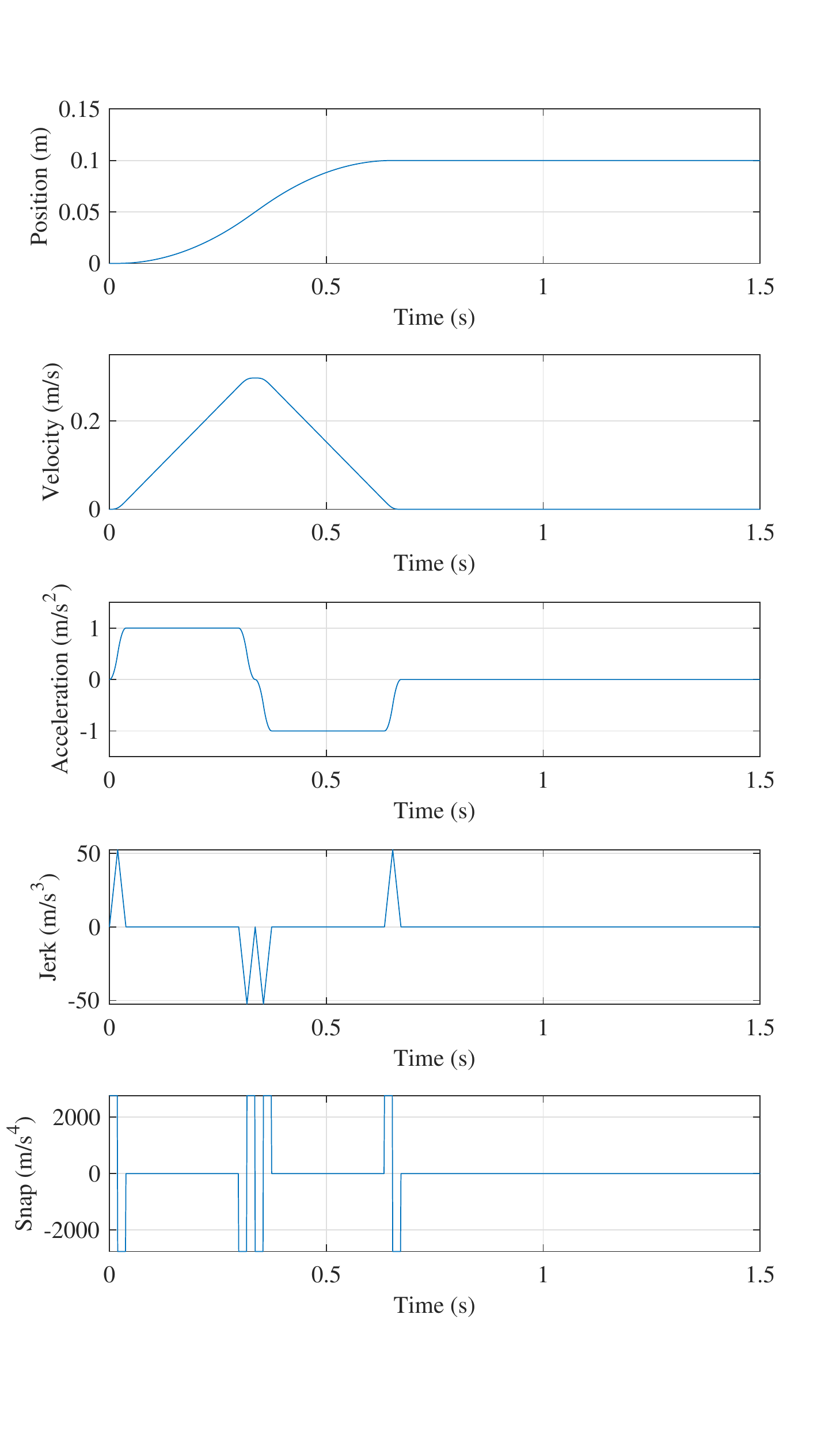}}
	\caption{Reference profile used in the simulation.}
	\label{fig.reference}
\end{figure}

\subsection{Convergence Analysis}

In this subsection, the convergence of the proposed method is proved.  

\noindent\textbf{Theorem 1} For the general uncertain mechatronic system ($\ref{regular-form}$) with control laws ($\ref{u}$)-($\ref{ST2}$) and ($\ref{ilc}$), $e_i$ and $s_i$ will be asymptotically converging to 0 as the iteration $i$ increasing.

\noindent\textbf{Proof of Theorem 1} Here, a Lyapunov function is designed to prove the convergence of $e_i$ and $s_i$. 
\begin{equation}
V_i=V_i^1+V_i^2 
\label{V}
\end{equation}
\noindent with
\begin{IEEEeqnarray}{rcl}
	V_i^1 &=& \frac{1}{2}\beta\varphi_i^2  \\
	V_i^2 &=& \frac{1}{2}\int_{0}^{t}q^{-1}\tilde{\Psi}_i^T\tilde{\Psi}_i\:d\tau  
\end{IEEEeqnarray}
First of all, the difference of $V_i^1$ between the $(i+1)$-th and the $i$-th iterations is written as
\begin{equation} 
\begin{aligned} \triangle V_{i+1}^1 =& V_{i+1}^1-V_{i}^1 \\                       
=& \frac{1}{2}\beta\varphi_{i+1}^2-\frac{1}{2}\beta\varphi_i^2
\label{V1_1}
\end{aligned}       
\end{equation}
From ($\ref{varrho}$), we have 
\begin{equation} 
\begin{aligned} 
\frac{d}{dt}\bigg(\frac{1}{2}\varphi_{i+1}^2\bigg) &= \dot{\varphi}_{i+1}(t)\varphi_{i+1}(t)\\             &= \frac{d}{ds_{i+1}}\varphi_{i+1}\cdot\dot{s}_{i+1}\cdot\varphi_{i+1} \\                                            
&= \varrho_{i+1}\dot{s}_{i+1}
\label{V1_2}
\end{aligned}       
\end{equation}
Substitute ($\ref{ST1}$), $\eqref{ST2}$, ($\ref{sigma-dynamics3}$), and ($\ref{V1_2}$) into ($\ref{V1_1}$), we have
\begin{equation} 
\begin{aligned} 
&\triangle V_{i+1}^1\\
=& \:\beta\int_{0}^{t}\varrho_{i+1}\tilde{\Psi}_{i+1}\:d\tau-\frac{k_1\beta}{2}\int_{0}^{t}\vert s_{i+1}\vert^{\frac{1}{2}}\bigg({\rm{sgn}}\big(s_{i+1}\big)\bigg)^2\:d\tau \\ 
&-\frac{3}{2}k_1k_c\beta\int_{0}^{t}\bigg(\vert s_{i+1}\vert^{\frac{1}{2}}{\rm{sgn}}\big(s_{i+1}\big)\bigg)^2\:d\tau \\ 
&-k_1\beta k_c^2\int_{0}^{t}\vert s_{i+1}\vert^{\frac{3}{2}}\:d\tau  
-\frac{k_1\beta}{2}k_c\int_{0}^{t}\vert s_{i+1}\vert\:d\tau   \\       
&-\frac{3}{2}k_1k_c^2\beta\int_{0}^{t}\vert s_{i+1}\vert^{\frac{3}{2}}\:d\tau        
-k_1k_c^3\beta\int_{0}^{t}\big(s_{i+1}\big)^2\:d\tau   \\  
&-\frac{k_2\beta}{2}\cdot\bigg(\int_{0}^{t}\varrho_{i+1}\:d\tau\bigg)^2-\frac{1}{2}\beta\varphi_i^2
\label{V1_3}
\end{aligned}      
\end{equation}
Then, the difference of $V_i^2(t)$ between the $(i+1)$-th and the $i$-th iterations is written as
\begin{equation} 
\begin{aligned} 
&\triangle V_{i+1}^2 = V_{i+1}^2-V_{i}^2 \\                                             
=& -\frac{1}{q}\int_{0}^{t}\big((\hat{\Psi}_{i+1}-\hat{\Psi}_i\big)^T\big(\Psi_{i+1}-\hat{\Psi}_{i+1}\big)\:d\tau \\
&-\frac{1}{2q}\int_{0}^{t}\big((\hat{\Psi}_{i+1}-\hat{\Psi}_i\big)^T\big(\hat{\Psi}_{i+1}-\hat{\Psi}_i\big)\:d\tau 
\label{V2_1}
\end{aligned}       
\end{equation}
Substitute ($\ref{ilc}$) into ($\ref{V2_1}$), we have
\begin{equation} 
\begin{aligned} 
\triangle V_{i+1}^2 =& -\beta\int_{0}^{t}\varrho_{i+1}\tilde{\Psi}_{i+1}\:d\tau-2q\int_{0}^{t}(\beta\:\varrho_{i+1}\big)^2\:d\tau 
\label{V2_2}
\end{aligned}       
\end{equation}
Finally, combine ($\ref{V1_3}$), ($\ref{V2_2}$) with ($\ref{V}$), we get
\begin{equation} 
\begin{aligned} 
&\triangle V_{i+1} = \triangle V_{i+1}^1+\triangle V_{i+1}^2 \\             
=& \:\beta\int_{0}^{t}\varrho_{i+1}\tilde{\Psi}_{i+1}\:d\tau \\&-\frac{k_1\beta}{2}\int_{0}^{t}\vert s_{i+1}\vert^{\frac{1}{2}}\bigg({\rm{sgn}}\big(s_{i+1}\big)\bigg)^2\:d\tau \\ 
&-\frac{3}{2}k_1k_c\beta\int_{0}^{t}\bigg(\vert s_{i+1}(\tau)\vert^{\frac{1}{2}}{\rm{sgn}}\big(s_{i+1}(\tau)\big)\bigg)^2\:d\tau \\ 
&-k_1\beta k_c^2\int_{0}^{t}\vert s_{i+1}(\tau)\vert^{\frac{3}{2}}\:d\tau \\&-\frac{k_1\beta}{2}k_c\int_{0}^{t}\vert s_{i+1}(t)\vert\:d\tau\\       
&-\frac{3}{2}k_1k_c^2\beta\int_{0}^{t}\vert s_{i+1}(t)\vert^{\frac{3}{2}}\:d\tau\\&-k_1k_c^3\beta\int_{0}^{t}\big(s_{i+1}(t)\big)^2\:d\tau   \\  
&-\frac{k_2\beta}{2}\cdot\bigg(\int_{0}^{t}\varrho_{i+1}(\tau)\:d\tau\bigg)^2-\frac{1}{2}\beta\varphi_i(t)^2 \\
&-\beta\int_{0}^{t}\varrho_{i+1}(\tau)\tilde{\Psi}_{i+1}(\tau)\:d\tau-2q\int_{0}^{t}(\beta\:\varrho_{i+1}(\tau)\big)^2\:d\tau\\ \leq&\: 0 \label{V-1}
\end{aligned}     
\end{equation}
It indicates that the Lyapunov function $V_i$ is non-negative and monotonically decreasing in the iteration domain. Thus, $\varphi_i$ is convergent, and thus the convergence of $s_i$ will be consequently validated. As $\lambda>0$, the convergence ability of $e_i$ is also guaranteed. \hfill{$\Box$}

\begin{table}\centering	
	\caption{System parameters of the dual-drive H-gantry stage.}	
	\label{nomenclature}       
	\begin{tabular}{llll}		
		\hline\noalign{\smallskip}		
		Symbol & Description & Value & Unit  \\		
		\noalign{\smallskip}\hline\noalign{\smallskip}		
		$m_e$ & Mass of end-effector & 11.512 & kg \\		
		$m_c$ & Mass of cross-arm & 4.371 & kg \\
		$m_1$ & Mass of X1 carriage & 1.728 & kg \\		
		$m_2$ & Mass of X2 carriage & 1.586 & kg  \\	
		$K_v$ & Stiffness of flexure & 8693.7 & N/m \\
		$\Gamma_1$ & Damping coefficient in X1 carriage & 172.7 & N$\cdot$s/m \\		
		$\Gamma_2$ & Damping coefficient in X2 carriage & 172.7 & N$\cdot$s/m  \\	
		$\Gamma_e$ & Damping coefficient in Y carriage & 172.7 & N$\cdot$s/m \\
		$K_{f}$ & Force constant & 100 & N/A \\
		\noalign{\smallskip}\hline		
	\end{tabular}	
\end{table}

\begin{table}\centering	
	\caption{Indices for convergence analysis.}	
	\label{index}       
	\begin{tabular}{ll}		
		\hline\noalign{\smallskip}		
		Symbol & Description  \\		
		\noalign{\smallskip}\hline\noalign{\smallskip}		
		RMSE & root-mean-square value for trajectory error \\		
		RMSE-$e_x$ & root-mean-square value for X-axis tracking error  \\
		RMSE-$e_y$ & root-mean-square value for Y-axis tracking error  \\		
		MaxAE & maximum absolute value for trajectory error \\		
		MaxAE-$e_x$ & maximum absolute value for X-axis tracking error  \\
		MaxAE-$e_y$ & maximum absolute value for Y-axis tracking error  \\
		RMSSV-$s_1$ & root-mean-square value for X1-carriage sliding variable \\		
		RMSSV-$s_2$ & root-mean-square value for X2-carriage sliding variable  \\
		RMSSV-$s_y$ & root-mean-square value for Y-carriage sliding variable  \\
		MaxASV-$s_1$ & maximum absolute value for X1-carriage sliding variable \\		
		MaxASV-$s_2$ & maximum absolute value for X2-carriage sliding variable  \\
		MaxASV-$s_y$ & maximum absolute value for Y-carriage sliding variable  \\
		\noalign{\smallskip}\hline		
	\end{tabular}	
\end{table}

\begin{figure}[t]
	\centerline{\includegraphics[trim = 0cm 0cm 0cm 0cm, width=8.5cm]{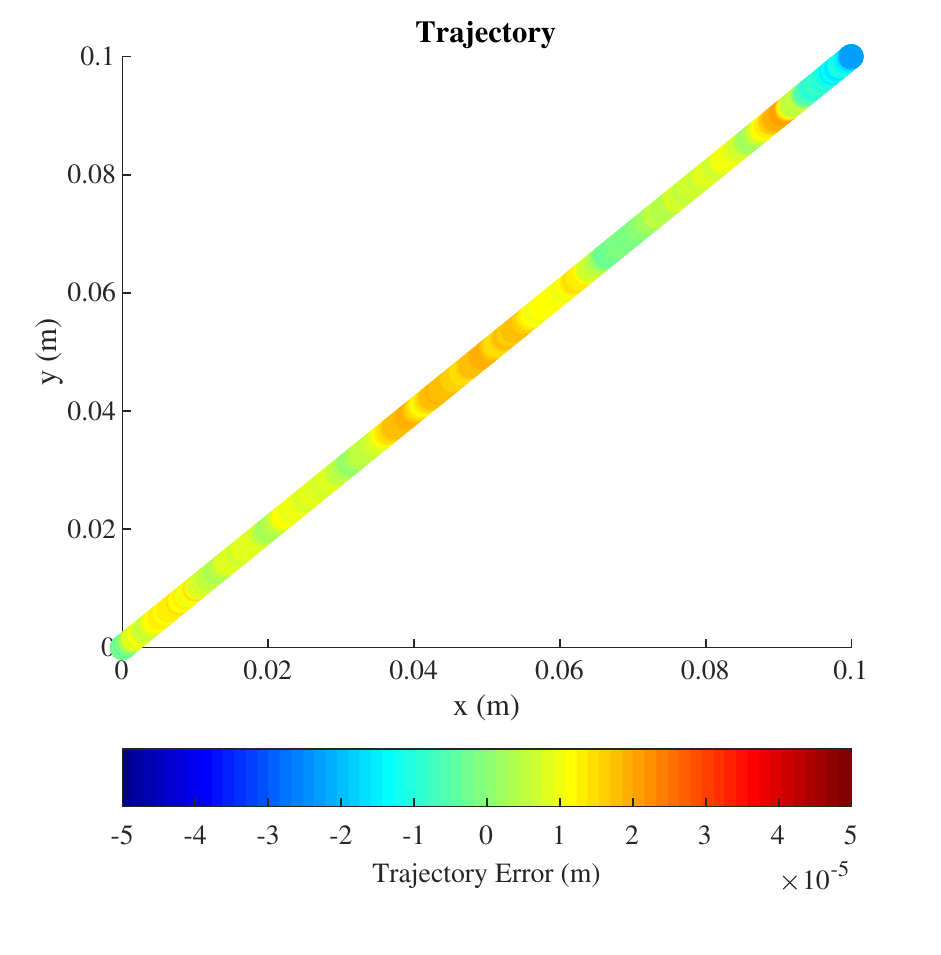}}
	\caption{Trajectory error in the 6th iteration.}
	\label{fig.trajectory}
\end{figure}

\begin{figure}[t]
	\centerline{\includegraphics[trim = 0cm 0cm 0cm 1cm, width=9cm]{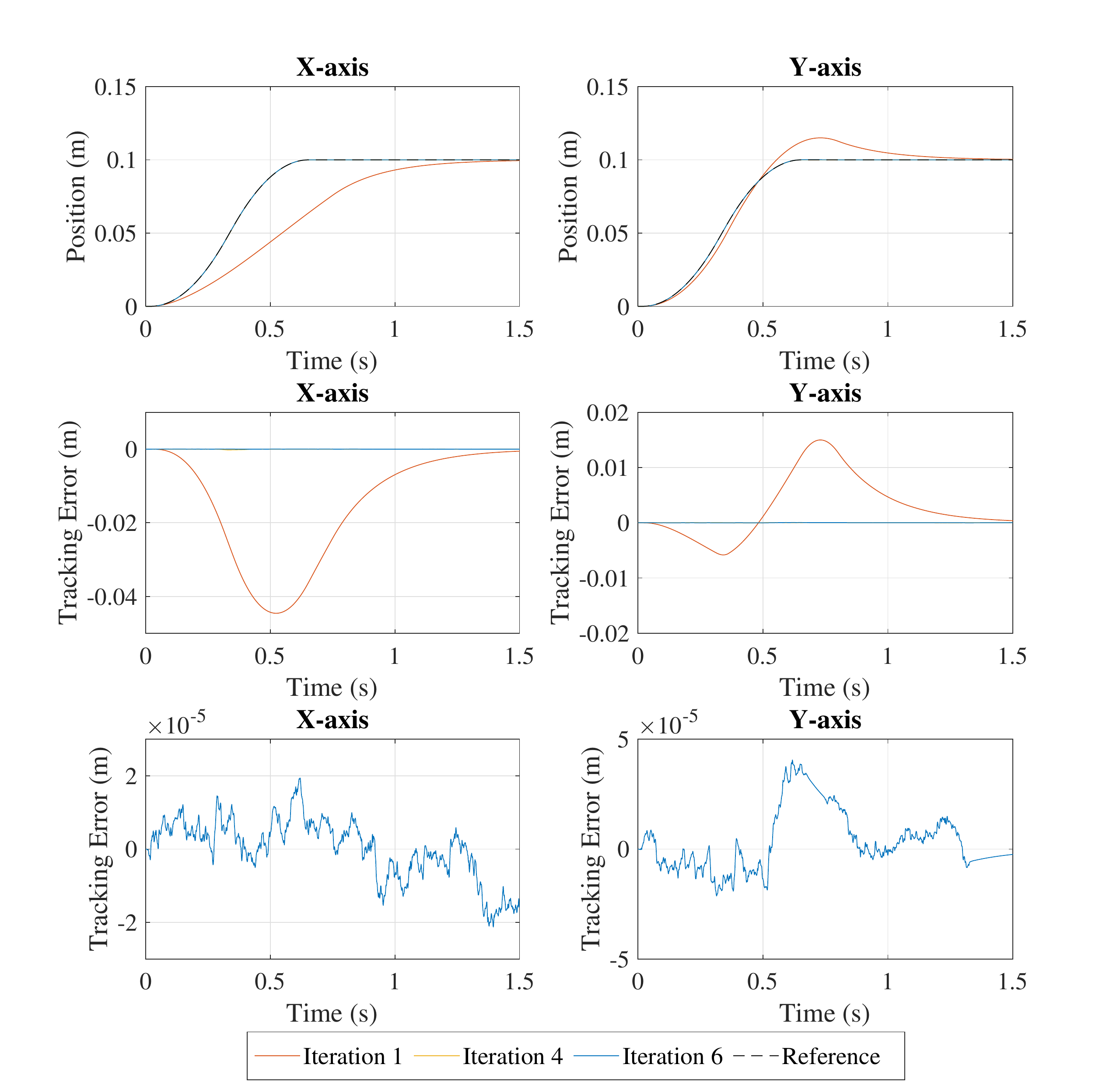}}
	\caption{\textbf{Tracking performance in the 1st, 4th, and 6th iteration.} \textbf{Top-Left.} Position performance (X-axis). \textbf{Top-Right.} Position performance (Y-axis). \textbf{Center-Left.} Tracking error (X-axis). \textbf{Center-Right.} Tracking error (Y-axis). \textbf{Bottom-Left.} Tracking error in the 6th iteration (X-axis). \textbf{Bottom-Right.} Tracking error in the 6th iteration (Y-axis).}
	\label{fig.e}
\end{figure}

\begin{figure}[t]
	\centerline{\includegraphics[trim = 0cm 0cm 0cm 1cm, width=9cm]{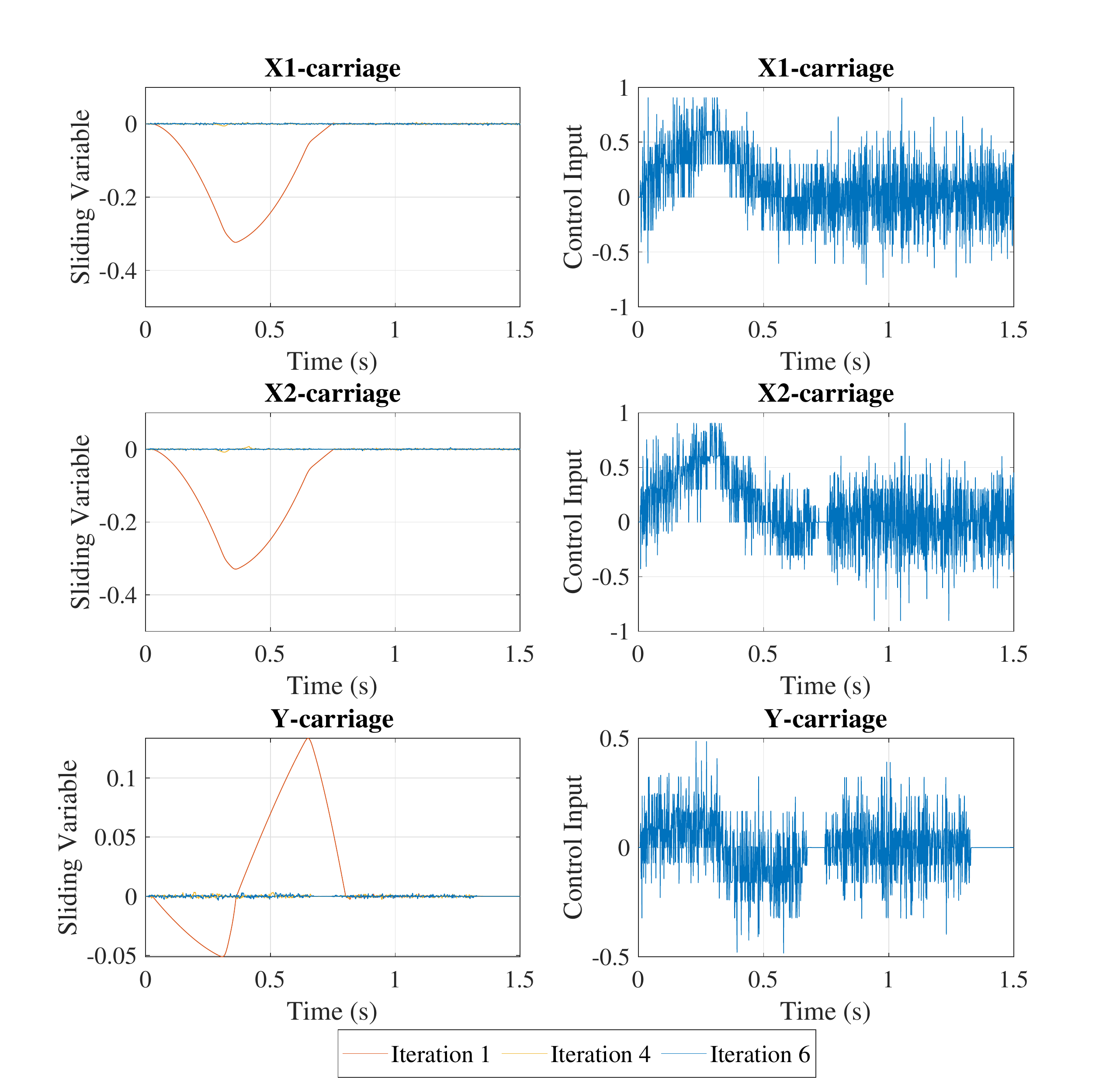}}
	\caption{\textbf{Sliding variables and control inputs of the three carriages in the 1st, 4th, and 6th iteration.} \textbf{Top-Left.} Sliding variables (X1-carriage). \textbf{Top-Right.} Control input (X1-carriage). \textbf{Center-Left.} Sliding variables (X2-carriage). \textbf{Center-Right.} Control input (X2-carriage). \textbf{Bottom-Left.} Sliding variables (Y-carriage). \textbf{Bottom-Right.} Control input (Y-carriage).}
	\label{fig.s}
\end{figure}

\begin{figure}[t]
	\centerline{\includegraphics[trim = 0cm 1cm 0cm 0.8cm, width=8.5cm]{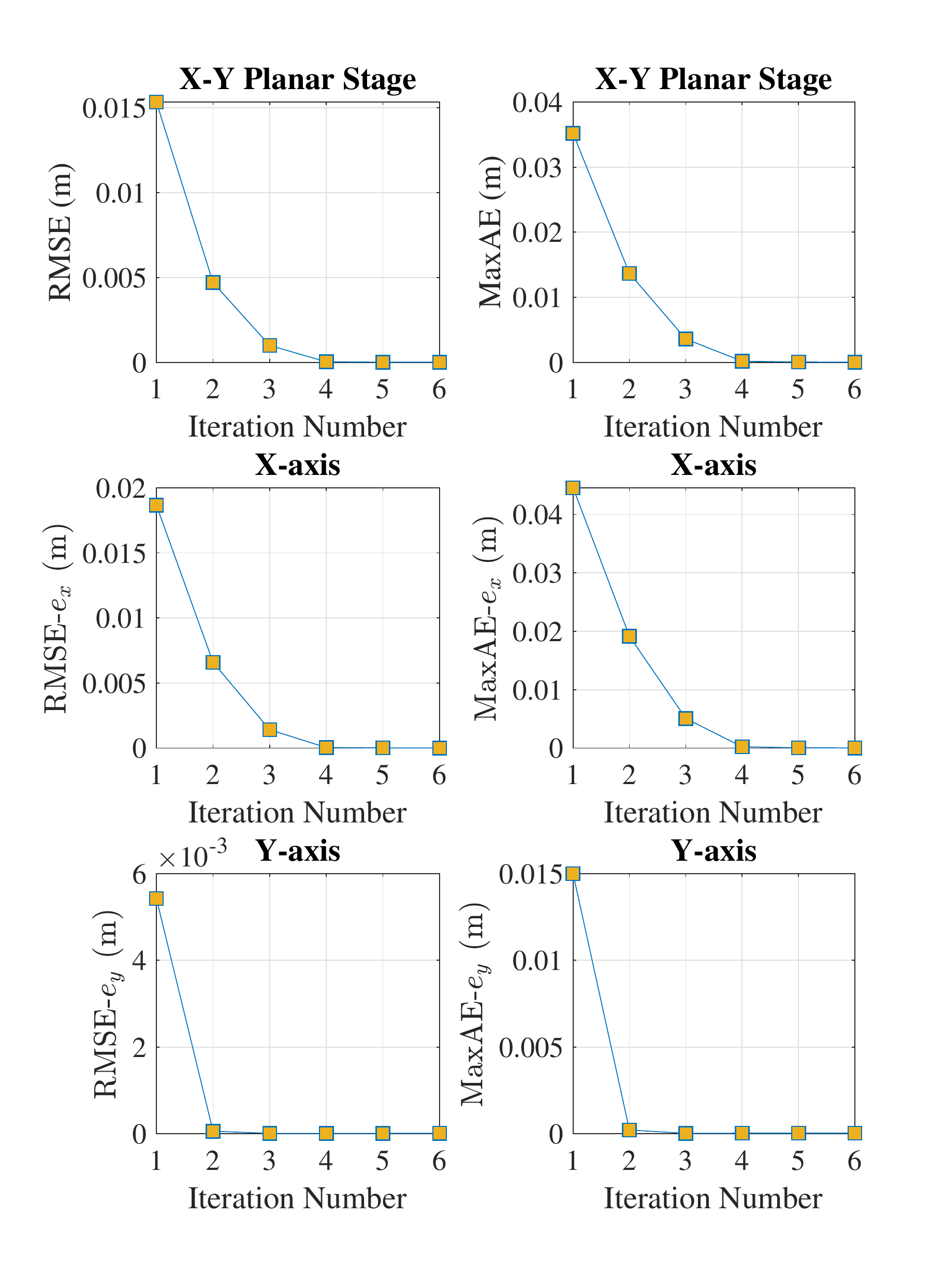}}
	\caption{\textbf{Convergence analysis of trajectory error and tracking error.} \textbf{Top-Left.} RMSE of Trajectory error. \textbf{Top-Right.} MaxAE of Trajectory error. \textbf{Center-Left.} RMSE of X-axis tracking error. \textbf{Center-Right.} MaxAE of X-axis tracking error. \textbf{Bottom-Left.} RMSE of Y-axis tracking error. \textbf{Bottom-Right.} MaxAE of Y-axis tracking error.}
	\label{fig.e_converge}
\end{figure}

\begin{figure}[t]
	\centerline{\includegraphics[trim = 0cm 1cm 0cm 0.8cm, width=8.5cm]{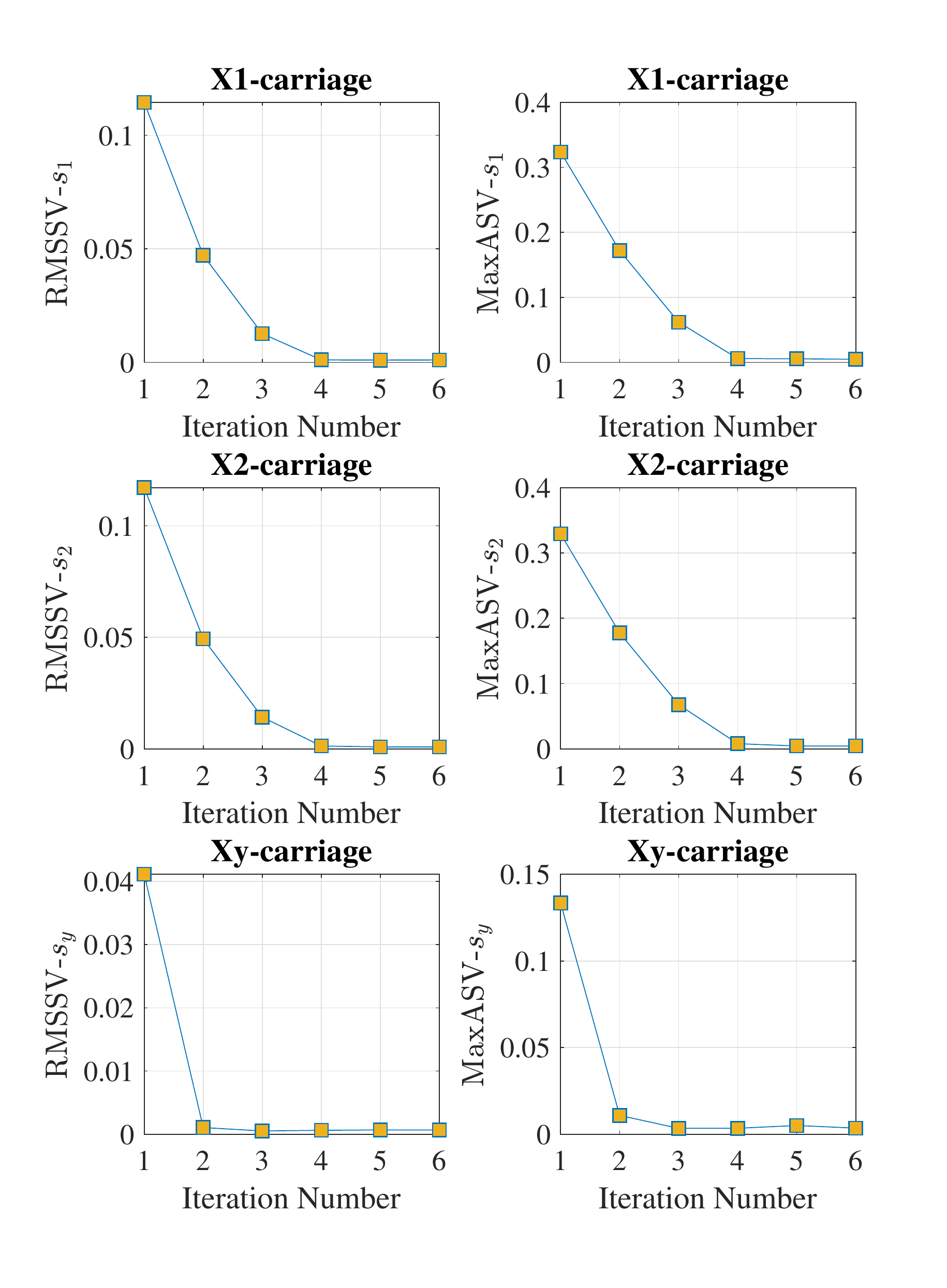}}
	\caption{\textbf{Convergence analysis of sliding variables.} \textbf{Top-Left.} RMSSV of X1-carriage sliding variable. \textbf{Top-Right.} MaxASV of X1-carriage sliding variable. \textbf{Center-Left.} RMSSV of X2-carriage sliding variable. \textbf{Center-Right.} MaxASV of X2-carriage sliding variable. \textbf{Bottom-Left.} RMSSV of Y-carriage sliding variable. \textbf{Bottom-Right.} MaxASV of Y-carriage sliding variable.}
	\label{fig.s_converge}
\end{figure}

\section{Simulation Analysis}

In this section, a case study on a flexure-joint dual-drive H-gantry stage is presented and analyzed. Fig. \ref{fig.gantry} is the mechatronic system in this case study. The detailed modeling of this system can be found in~\cite{kamaldin2018novel}. According to~\cite{ma2017integrated,kamaldin2018novel}, the model can be simplified as
\begin{IEEEeqnarray}{rcl}
	M_1\ddot{y}_1 &=& K_{f}u_1-\Gamma_1\dot{y}_1+v+d_1  \\
	M_2\ddot{y}_2 &=& K_{f}u_2-\Gamma_2\dot{y}_2-v+d_2  \\
	M_e\ddot{y}_y &=& K_{f}u_y-\Gamma_e/\cos(\Theta)\dot{y}_y+d_y
\end{IEEEeqnarray}
\noindent where $y_1$, $y_2$, $y_y$ denote the position of X1-carriage, X2-carriage, Y-carriage, respectively. $u_1$, $u_2$, $u_y$ denote the control input of the three carriages. $M_1=(m_e+m_c)/2+m_1$, $M_2=(m_e+m_c)/2+m_2$, and $M_e=m_e/\cos(\Theta)$. $\Theta$ is the rotation angle of the cross arm. $d_1$, $d_2$, and $d_y$ are lumped disturbance (including friction) of X1 carriage, X2 carriage, and Y carriage, respectively. $v=K_v(y_1-y_2)$ is the coupling force between the two parallel carriages. The system parameters used in this simulation part are listed in Table \ref{nomenclature}. 

In this case study, the motion dynamics function for each axis can be written in the form of \eqref{regular-form} and the multi-axis cooperation problem is transformed to model-free robust control problem for each axis. Here, a planar point-to-point task is designed to validate the effectiveness of the proposed method. Each carriage is tracked same S-shaped reference profile, which is shown in Fig. $\ref{fig.reference}$. Additionally, the iterative learning law ($\ref{ilc}$) contains signum term, which will cause input chattering when the parameter $q$ is set too large. Considering this issue, a modification \cite{armstrong2019improved} is used when calculating the control input in iterations, which is given by
\begin{equation}
u^{ilc}_{i+1}(k)=\left\{\begin{array}{ll} 
u^{ilc}_i(k)-2q\beta\varrho_{i+1}(k)  &{\rm{if }} \: \vert s_{i+1}(k)\vert>\epsilon \\
u^{ilc}_{i+1}(k-1) & {\rm{if }} \: \vert s_{i+1}(k)\vert\leq\epsilon                       
\end{array}\label{modify}  \right.
\end{equation}
To analyze the simulation results, not only tracking error of each axis but also trajectory error is considered. In detail, the three carriages (X1, X2, and Y) are given the same S-curve reference signal so that the trajectory in the X-Y stage should be a point-to-point line ($y=x$). Thus, we define trajectory error as the distance between actual positions and the reference trajectory $e_d=|x-y|/\sqrt{2}$, where $x=(y_1+y_2)/2$ is the position of the end-effector in X-axis, $y=y_y$ is the position of the end-effector in Y-axis. Also notice that, $s_j$ denotes the sliding variable of the $j$ carriage ($j=1,2,y$) and $e_k$ denotes the tracking error in each axis ($k=x,y$). Here, also mention that, the tracking error of X-axis is defined as $e_x=(y_1+y_2)/2-r$ and the tracking error of Y-axis is defined as $e_y=y_y-r$. Moreover, to show the convergence ability, some indices are defined for convenience which are shown in Table \ref{index}.

In this case study, the super-twisting gains are set as $k_1=0.1$, $k_2=0.1$, and the linear term gain are set as $k_c=0.1$. The iterative learning parameters $q$ and $\beta$ are set as $q=1.5$ and $\beta=0.1$. The positive parameter of switching function is set as $\lambda=5$. At first, the trajectory error is shown in Fig. $\ref{fig.trajectory}$ where the color represents the trajectory accuracy as shown in the color bar. It is obvious seen that the trajectory error fluctuates within the scale of $-2\times10^{-5}$m to $2\times10^{-5}$m. Then, the sliding variables and control inputs are shown in Fig. $\ref{fig.s}$, which indicates that the chattering is attenuated effectively. In terms of the convergence ability, Fig. $\ref{fig.e_converge}$ and Fig. $\ref{fig.s_converge}$ represent root-mean-square values and maximum absolute values of tracking errors and sliding variables. We can see that the convergence is completed in 4 iterations.

\section{Conclusion}

This paper presents a generalized iterative super-twisting method, which is a totally model-free controller to improve motion performance for uncertain mechatronic systems. In addition, the convergence ability of the proposed method is analyzed with rigorous proof. Moreover, a simulation case study on the flexure-joint dual-drive H-gantry stage is given. The results indicate that the tracking performance and sliding accuracy are highly improved as well as the chattering and uncertainties effects are suppressed.

\bibliographystyle{IEEEtran}
\bibliography{IEEEabrv,Ref}

\end{document}